\documentclass[aps,twocolumn,floats,superscriptaddress,prd,nofootinbib]{revtex4-1}
\newlength{\picwidth}
\usepackage{graphicx,epsfig,amsmath,amssymb,amsfonts,enumerate,bm}
\usepackage{color}
\usepackage{hyperref}

\def\be{\begin{equation}}
\def\ee{\end{equation}}
\def\ba{\begin{eqnarray}}
\def\ea{\end{eqnarray}}

\begin{document}

%%%%%%%%%%%%%%%%%%%%%%%%%%%%%%%%%%%%%%%%%%%%%%%%%%%%%%%%
%%%%%%%%%%%%%%%%%%%%%%%%%%%%%%%%%%%%%%%%%%%%%%%%%%%%%%%%

\title{Neutrino physics from future weak lensing surveys}

\author{R. Ali Vanderveld}
\affiliation{Kavli Institute for Cosmological Physics, Enrico Fermi Institute, University of Chicago, Chicago, IL 60637}
\author{Wayne Hu}
\affiliation{Kavli Institute for Cosmological Physics, Enrico Fermi Institute, University of Chicago, Chicago, IL 60637}
\affiliation{Department of Astronomy and Astrophysics, University of Chicago, Chicago, IL 60637}
\date{\today}

%%%%%%%%%%%%%%%%%%%%%%%%%%%%%%%%%%%%%%%%%%%%%%%%%%%%%%%%
\begin{abstract}

Given recent indications of additional neutrino species and cosmologically significant neutrino masses, we analyze their signatures in the  weak lensing shear power spectrum.   We find that a shear deficit in the  20-40\% range or excess in the 20-60\% range cannot be explained by variations in parameters of the flat $\Lambda$CDM model that are allowed by current observations of the expansion history from Type Ia supernovae, baryon acoustic oscillations, and local measures of the Hubble constant $H_0$, coupled with observations of the cosmic microwave background from WMAP9 and the SPT 2500 square degree survey. Hence such a shear deficit or excess would indicate large masses or extra species, respectively, and we find this to be independent of the flatness assumption.  We also discuss the robustness of these predictions to cosmic acceleration physics and the means by which shear degeneracies in joint variation of mass and species can be broken.

\end{abstract}

\maketitle

%%%%%%%%%%%%%%%%%%%%%%%%%%%%%%%%%%%%%%%%%%%%%%%%%%%%%%%%
%%%%%%%%%%%%%%%%%%%%%%%%%%%%%%%%%%%%%%%%%%%%%%%%%%%%%%%%
\section{Introduction}
\label{sec:introduction}

As our cosmological observations become ever more precise, our ability to probe smaller scales continues to advance, thereby allowing us to study the physics of structure formation beyond the standard cold dark matter paradigm.
% and into the regime where baryons, warm dark matter, and neutrinos have a significant effect. 
In particular, we are now able to use cosmology to learn about neutrino properties, including the sum of their masses $M_{\nu}$ and the effective number of species $N_{\rm{eff}}$, both of which imprint their signatures on the small-scale matter power spectrum. Massive neutrinos act as hot or warm dark matter, thereby suppressing structure formation below their thermal free-streaming scale, while adding or subtracting relativistic species changes the ratio of the acoustic and damping angular scales of the cosmic microwave background (CMB) \cite{Hu:1996vq,Hou:2011ec}. The former is currently constrained to be $M_{\nu} \gtrsim 0.05~\rm{eV}$ by solar, atmospheric, and laboratory experiments \cite{BeckerSzendy:1992hq,Fukuda:1998fd,Ahmed:2003kj}, and (roughly) $M_{\nu} \lesssim 0.6~\rm{eV}$ from cosmology \cite{Tereno:2008mm,Reid:2009nq,Thomas:2009ae,Mantz:2009rj}.

As for $N_{\rm{eff}}$, recent  oscillation and reactor experiments
\cite{AguilarArevalo:2010wv,Mention:2011rk} lend support to the sterile
neutrino interpretation of the LSND electron antineutrino appearance result
\cite{Aguilar:2001ty} and $M_{\nu} \gtrsim 0.4$ eV while other neutrino results inhibit a simple global 
explanation (see e.g.~Ref.~\cite{Abazajian:2012ys} for a recent review).  Meanwhile,
recent observations of the CMB damping tail \cite{Dunkley:2010ge,Keisler:2011aw,Story:2012wx,Hou:2012xq}, Sunyaev Zel'dovich-selected clusters \cite{Benson:2011ut},
and baryon acoustic oscillations (BAO) \cite{Anderson:2012sa} provide further hints of extra neutrino species.

In this paper we explore how weak gravitational lensing fits into this picture, in light of forthcoming lensing-optimized large-area surveys such as with the ground-based Dark Energy Survey (DES) \cite{des_url} and Large Synoptic Survey Telescope (LSST) \cite{lsst_url}, from the balloon-borne High Altitude Lensing Observatory \cite{Rhodes:2012sr}, or from space with {\it Euclid} \cite{euclid_url} and the Wide-Field Infrared Survey Telescope (\!{\emph{WFIRST}}) \cite{wfirst_url}. Weak lensing, whereby the images of distant galaxies are distorted by the gravitational field of matter in the foreground, can be a powerful cosmological probe provided that we have sufficient systematics control. By extracting the weak lensing shear and its evolution with redshift we are able to robustly map out the gravitational potential of the Universe and how it changes with time. The power spectrum of this ``cosmic shear" is directly related to the underlying matter power spectrum. Despite promising results, e.g.~\cite{Fu:2007qq,Kilbinger:2008gk,Massey:2007gh,Massey:2007wb,Schrabback:2009ba} 
(see Refs.~\cite{Takada:2003ef,Hoekstra:2008db,Massey:2010hh} for reviews), the current constraining power of cosmic shear is very limited. Since we do not know the intrinsic shapes of individual galaxies, we must average them over finite patches of sky, making weak lensing a necessarily statistical measure whose constraining power is directly related to sky coverage \cite{Hu:1998az,Amara:2006kp}. Future data sets will be optimized in this respect, but in the meantime it is particularly timely to determine our expectations.

We can robustly test any given cosmological model class by exploiting consistency relations between observables pertaining to the expansion history and those pertaining to structure growth \cite{Mortonson:2008qy,Mortonson:2009hk,Mortonson:2010mj,Vanderveld:2012ec}. Given one, coupled with a class of cosmological models with tunable parameters, we can predict the other and then compare our predictions to data. If the data points lie significantly outside of the prediction contours, then the model class in question is falsified. For instance, just one cluster that is massive and at high-enough redshift could falsify all $\Lambda$CDM and quintessence models if its mass and redshift fall significantly outside of what we predict based on Type Ia supernovae (SNe), BAO, local measurement of the Hubble constant ($H_0$), and the CMB \cite{Mortonson:2010mj}. In this way we can take advantage of the wealth of data already in hand from the CMB and distance measures to predict what we expect for these anticipated future weak lensing observations. Our analysis here builds upon \cite{Vanderveld:2012ec} to explore the effects of neutrinos.

In what follows we will add neutrinos to this prediction framework, to explore how their masses and number of species change weak lensing observables.  
We find that, for a fixed CMB, these two properties shift the cosmic shear power spectrum in opposite directions; adding relativistic species amplifies the shear power, whereas endowing the neutrinos with nonzero masses reduces power. Given that quintessence can only decrease the amount of power \cite{Vanderveld:2012ec}, the former
provides qualitatively distinct predictions.

This paper is organized as follows. We review our methodology in \S \ref{sec:methodology}, including the data sets we use, our Markov-Chain Monte Carlo (MCMC) analysis, and the calculation of posterior probability distributions for cosmic shear observables. We then discuss our results in \S \ref{sec:results}, including predictions for four cases -- three massless neutrinos, three massive neutrinos, a variable number of massless neutrinos, and both variable mass and number. We conclude in \S \ref{sec:discussion}.

%%%%%%%%%%%%%%%%%%%%%%%%%%%%%%%%%%%%%%%%%%%%%%%%%%%%%%%%
\section{Methodology}
\label{sec:methodology}

We  describe here the data sets we use and our procedure for predicting the cosmic shear power spectrum under various assumptions for the total neutrino mass $M_{\nu}$ and effective number of species $N_{\rm{eff}}$.  Our methodology is similar to that of Refs.~\cite{Mortonson:2008qy,Mortonson:2009hk,Mortonson:2010mj,Vanderveld:2012ec}.

\subsection{Data sets}
\label{sec:data}

We use the redshifts, luminosity distances, and systematic uncertainty estimates of the Union2 Type Ia SN sample \cite{Amanullah:2010vv}. This sample includes 557 SNe out to a redshift $z=1.12$, where all light curves have been uniformly reanalyzed using the SALT2 fitter \cite{Guy:2007dv}.

We use measurements of the BAO feature from Ref.~\cite{Percival:2009xn} (which includes data from SDSS and the 2-degree Field Galaxy Redshift Survey), the WiggleZ Dark Energy Survey \cite{Blake:2011en}, and the SDSS-III Baryon Oscillation Spectroscopic Survey (BOSS) \cite{Padmanabhan:2012hf,Anderson:2012sa}. These measurements extend out to $z=0.73$ and are reported as distances relative to the sound horizon, $D_V(z) /r_s$, where $D_V(z)\equiv [(1+z)^2 D_A^2(z)cz/H(z)]^{1/3}$, $D_A$ is the angular diameter distance, $H(z)$ is the Hubble expansion rate, and $r_s$ is the sound horizon at last scattering. Even though these data sets have some overlap in both area and redshift, we treat all three as independent due to the different bias and type of galaxies that are targeted in each sample.

Unlike Refs.~\cite{Mortonson:2008qy,Mortonson:2009hk,Mortonson:2010mj,Vanderveld:2012ec}, we use CMB results from the most recent, 9-year release from the {\it WMAP} satellite (WMAP9) \cite{Hinshaw:2012fq}, computing the CMB angular power spectra using the code CAMB \cite{Lewis:1999bs,camb_url}. We now further add the publicly available 2500 square degree release of the SPT measurement of the CMB damping tail \cite{Story:2012wx} over the multipole range $650 < l < 3000$, as these smaller-scale peaks are sensitive to neutrino physics. Per Ref.~\cite{Story:2012wx}, we treat the SZ and point-source contributions as (three) additional nuisance parameters, choosing the same Gaussian priors for each.

Finally, we use the combined 
$H_0$ estimate from Ref.~\cite{Riess:2011yx}, of $H_0 = 73.8 \pm 2.4~{\rm km/s/Mpc}$. This measurement strengthens our constraint on $N_{\rm{eff}}$. 

\subsection{Parameter sets}
\label{sec:parameters}

We use this data set to probe flat $\Lambda$CDM models with varying $f_{\nu}$ and $N_{\rm{eff}}$. The former is the fraction of the dark matter density in the form of massive neutrinos
\be
f_{\nu}=\frac{\Omega_{\nu}}{\Omega_{\rm{DM}}}\,,
\ee
where
\be
\Omega_{\nu}=\frac{M_{\nu}}{93.14 h^2~{\rm eV}}
\ee
and $M_{\nu}$ is the sum of the neutrino masses. The parameter $N_{\rm{eff}}$ is the so-called effective number of neutrino species \cite{Mangano:2001iu}:
\be
\rho_{\rm{R}}=\left[1+\frac{7}{8}\left(\frac{4}{11}\right)^{4/3}N_{\rm{eff}}\right]\rho_{\gamma}\,,
\ee 
where $\rho_{\rm{R}}$ is the energy density in relativistic species and $\rho_{\gamma}$ is the energy density of photons. We increase (decrease) $N_{\rm{eff}}$ by adding (subtracting) massless species. The default is three species with degenerate masses. Unfortunately this means that $N_{\rm{eff}}<3.046$ leads to a negative number of massless species, an unphysical situation which is treated by CAMB as a negative energy density; we follow the SPT analyses \cite{Keisler:2011aw,Benson:2011ut,Story:2012wx,Hou:2012xq} and ignore this since $N_{\rm{eff}}>3.046$ is highly favored and indeed these unphysical cases make up no more than a few percent of the samples in our MCMC chains. We then set the primordial helium abundance $Y_p$ from the physical baryon density $\Omega_{\rm{b}}h^2$ by requiring ``big bang nucleosynthesis consistency" \cite{Simha:2008zj}, such that
\be
Y_p = 0.2485 + 0.0016 \left[ 273.9 \Omega_{\rm{b}}h^2 - 6 + 100\left(S-1\right)\right]\,,
\ee
where $S$ depends on the number of neutrino species
\be
S^2 = 1 + \frac{7}{43}\left(N_{\rm{eff}} - 3.046\right)\,.
\ee

The total parameter set we use is
\be
\bm{\theta} = \{\Omega_{\rm{b}}h^2, \Omega_{\rm{DM}}h^2, \tau, \theta_A, n_s, \ln A_s, f_{\nu}, N_{\rm{eff}}\}\,,
\label{eq:lcdmpar}
\ee
where $\Omega_{\rm{DM}}h^2$ is the present physical dark matter density relative to the critical density,  $\tau$ is the reionization optical depth, $\theta_A$ is the angular size of the acoustic scale at last scattering, $n_s$ is the spectral index of the power spectrum of initial fluctuations, and $A_s$ is the amplitude of the initial curvature power spectrum at $k_{\rm p}=0.05~{\rm Mpc}^{-1}$. All other parameters, including the Hubble constant $H_0 = 100\,h~{\rm km/s/Mpc}$, the present total matter density $\Omega_{\rm m}$, the dark energy density $\Omega_{\rm{DE}}$, and the amplitude of the matter power spectrum today $\sigma_8$, can be derived from this set. We will study four different cases: (1) $f_{\nu}=0$ and $N_{\rm{eff}}=3.046$, i.e.~standard flat $\Lambda$CDM, (2) $f_{\nu}$ is allowed to vary, $N_{\rm{eff}}=3.046$, (3) $N_{\rm{eff}}$ is allowed to vary, $f_{\nu}=0$, and (4) both $f_{\nu}$ and $N_{\rm{eff}}$ allowed to vary.

For a given set of parameters $\bm{\theta}$ that defines the cosmological model class in question, we use the CosmoMC code \cite{Lewis:2002ah,cosmomc_url} to sample from the joint posterior distribution,
\be
{\cal P}(\bm{\theta}|{\bf x})=\frac{{\cal L}({\bf x}|\bm{\theta}){\cal P}(\bm{\theta})}{\int d\bm{\theta}{\cal L}({\bf x}|\bm{\theta}){\cal P}(\bm{\theta})}\,,
\label{eq:bayes}
\ee
where ${\cal L}({\bf x}|\bm{\theta})$ is the likelihood of the dataset ${\bf x}$ given the model parameters $\bm{\theta}$ and ${\cal P}(\bm{\theta})$ is the prior probability density. For the standard $\Lambda$CDM parameters we use the same priors as in \cite{Vanderveld:2012ec} (flat priors that are wide enough to not limit our constraints), and we similarly choose wide priors for $f_{\nu}$ and $N_{\rm{eff}}$ that are informed by the current limits from data as summarized in Refs.~\cite{Benson:2011ut,Bird:2011rb}. In particular we choose the extremely conservative prior $1.047 < N_{\rm{eff}} < 10.0$, as in the SPT analyses.

\subsection{Weak lensing observables}
\label{sec:observables}

We can compute the posterior probability distribution for any derived statistic from the joint posterior distribution of the cosmological parameters. In particular, in order to compute the cosmic shear power spectrum, we must first compute the comoving angular diameter distance $D$ and the nonlinear matter power spectrum $\Delta^2_{\rm NL}$. In a flat universe (curvature $\Omega_{\rm{K}}=0$), the former is equal to the comoving radial coordinate and is related to the cosmological parameters through
\be
D(z) = \int_0^z\frac{dz'}{H(z')} \,.
\ee
Here the Hubble expansion rate is
\be
H(z)=H_0\left[\Omega_{\rm{m}}(1+z)^3+\left(1-\Omega_{\rm{m}}\right)\right]^{1/2}\,,
\ee
where the total matter density is
\be
\Omega_{\rm m}\equiv\Omega_{\rm DM}+\Omega_{\rm b}\,
\ee
and the contribution from radiation is assumed to be negligible.

We compute the $z=0$ linear matter power spectrum $\Delta_{\rm L}^2(k;0)$ using CAMB. The linear matter power spectrum at earlier redshifts then depends on the growth function of linear density perturbations. Massive neutrinos suppress growth in a scale-dependent manner, and we model this using the Eisenstein and Hu \cite{Eisenstein:1997jh} fitting function, 
\be
\Delta_{\rm L}^2(k;z) = \Delta_{\rm L}^2(k;0)\frac{T^2(k,z)}{T^2(k,0)}\frac{D_1^2(z)}{D_1^2(0)}\,,
\ee
where $T(k,z)$ and $D_1(z)$ are given by their Eqs.~(7) and (8), respectively, which we have modified according to Ref.~\cite{Kiakotou:2007pz} to improve accuracy in the case of three massive neutrinos. 
Note that $D_1(z)$ corresponds to the standard scale-independent growth function in the absence of neutrinos, and the scale-dependent effects of their free-streaming are encoded in $T(k,z)$.
Comparing to results from CAMB for nonzero redshifts, this fitting formula typically reproduces the growth to better than $1\%$ for all $k$ and $z$ we use here to compute the shear.

We compute the full nonlinear matter power spectrum at a given redshift using the Halofit fitting function \cite{Smith:2002dz} (see Ref.~\cite{Vanderveld:2012ec} for a summary), modified for the effects of massive neutrinos \cite{Bird:2011rb}. The original Halofit fitting functions have been found to only be accurate (even for the flat $\Lambda$CDM model) at up to the $5$--$10\%$ level compared with $N$-body results, for instance with the Coyote Universe project \cite{Heitmann:2008eq,Heitmann:2009cu,Lawrence:2009uk}.   We find that whether or not we use the massive neutrino modification
\cite{Bird:2011rb} 
leads to errors of order a few percent for the small neutrino masses considered here. These systematic errors are smaller than the statistical errors arising from our data sets in the same regime \cite{Vanderveld:2012ec}, and so we expect our results to be fairly robust to them. Likewise, for a wide range of baryonic effects, systematic shifts are at most comparable to current statistical errors \cite{Vanderveld:2012ec}.

The shear (or equivalently the convergence) power spectrum is then equal to
\be
{l^2 P_{\kappa} \over 2\pi} = {9\pi \over 4 c^4 l}  {\Omega_{\rm{m}}^2H_0^4  } \int_0^{\infty}dz {D^3 \over H} \frac{g^2(z)}{a^2}\Delta^2_{\rm NL}\left(\frac{l}{D};z\right)\,,
\label{shearpower}
\ee
where $k \approx l/D$ in units of Mpc$^{-1}$ in the Limber approximation and we have defined the geometric lensing efficiency factor
\be
g(z)\equiv\int^{\infty}_{z} dz' n(z')\frac{D'-D}{D'}\,.
\ee
The efficiency factor weights according to the source distribution in a given survey, $n(z)$, normalized such that $\int_0^{\infty}n(z) dz =1$. Here we use the model
\be
n(z)\propto  \left(\frac{z}{z_0}\right)^{\alpha}\exp\left[-\left(\frac{z}{z_0}\right)^{\beta}\right]\,,
\ee
with parameters $(z_0,\alpha,\beta)=(0.555,1.197,1.193)$ for a simplified model ground-based survey, such as CFHTLS or DES, with an approximate median redshift of $0.8$.

%%%%%%%%%%%%%%%%%%%%%%%%%%%%%%%%%%%%%%%%%%%%%%%%%%%%%%%%
\section{Results}
\label{sec:results}

%****************************************
\begin{figure*}
\includegraphics[width=\picwidth]{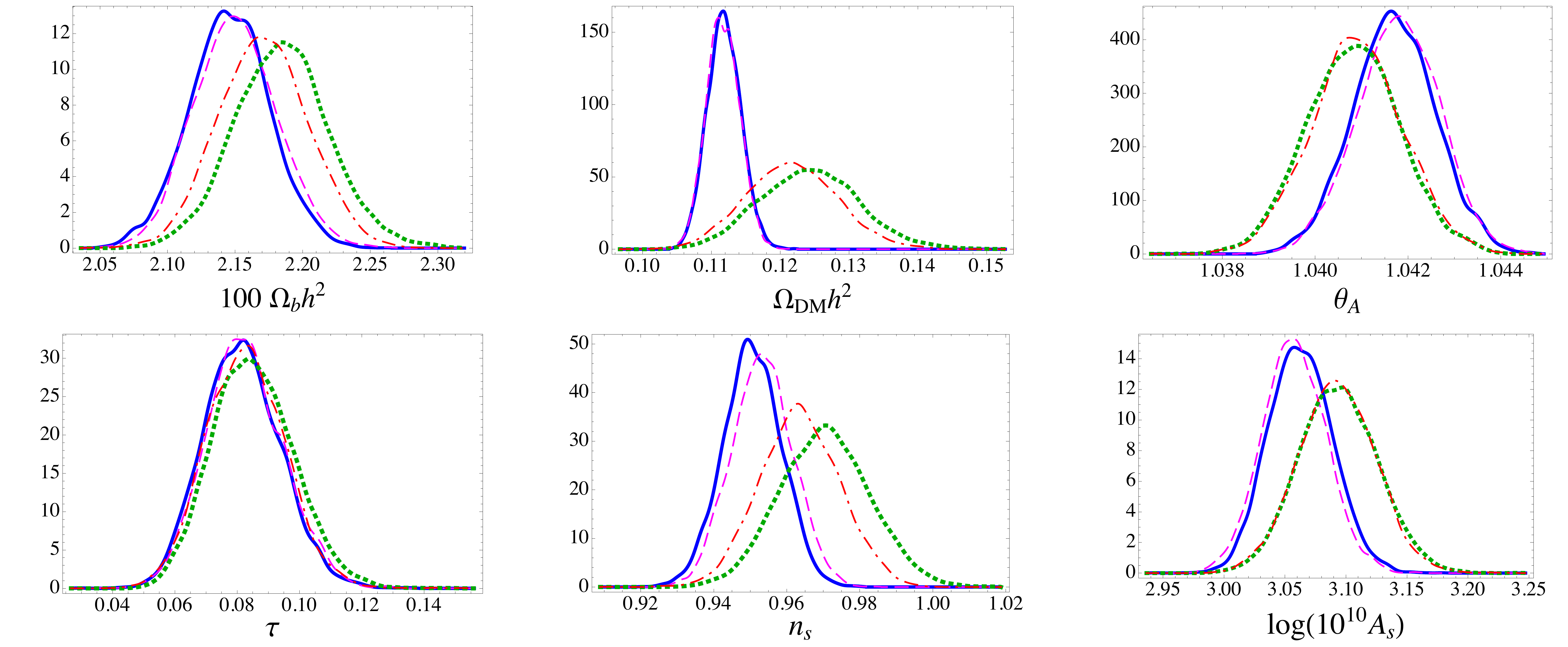}
\caption{One-dimensional constraints on the six cosmological parameters of our baseline flat $\Lambda$CDM model for our four cases: (1) three massless neutrinos (blue, thick solid); (2) $f_{\nu}$ allowed to vary (magenta, dashed); (3) $N_{\rm{eff}}$ allowed to vary (red, dot-dashed); and (4) both $f_{\nu}$ and $N_{\rm{eff}}$ allowed to vary (green, thick dotted). }
\label{fig:params} 
\end{figure*}
%****************************************

The histograms in Fig.~\ref{fig:params} illustrate what happens to our predictions for our six flat $\Lambda$CDM parameters $\{\Omega_{\rm{b}}h^2, \Omega_{\rm{DM}}h^2, \tau, \theta_A, n_s, \ln A_s, f_{\nu}, N_{\rm{eff}}\}$ when $f_{\nu}$ and/or $N_{\rm{eff}}$ are allowed to vary. All of our results from here on are presented with the following color-coding:
\begin{itemize}
\item[(1)] Blue: $f_{\nu}=0$ and $N_{\rm{eff}}=3.046$, i.e.~standard flat $\Lambda$CDM
\item[(2)] Magenta: $f_{\nu}$ is allowed to vary and $N_{\rm{eff}}=3.046$
\item[(3)] Red: $N_{\rm{eff}}$ is allowed to vary and $f_{\nu}=0$
\item[(4)] Green: both $f_{\nu}$ and $N_{\rm{eff}}$ are allowed to vary
\end{itemize}
We further summarize our constraints on $M_{\nu}$ and $N_{\rm{eff}}$ in Table~\ref{table:fnunnu} for each of these four cases. In the most general case (4), we constrain $M_{\nu} < 0.67$ eV at the 95\% confidence level and we find $N_{\rm{eff}} = 3.71\pm 0.35$. These constraints are in agreement with the current state-of-the-art as seen in the literature for cosmological probes, e.g.~Ref.~\cite{Giusarma:2012ph}. On the other hand, since the $M_{\nu}$ constraints are based on the CMB and
expansion history measurements rather than growth measurements (see e.g. \cite{Hu:1997mj}), they are 
less robust to generalizations of the flat $\Lambda$CDM model, e.g. the 
addition of spatial curvature (see below).

%****************************************
\begin{table}[h]
\caption{Constraints on the sum of the neutrino masses $M_{\nu}$ and effective number of species $N_{\rm{eff}}$ for our four cases: (1) three massless neutrinos; (2) $f_{\nu}$ allowed to vary; (3) $N_{\rm{eff}}$ allowed to vary; and (4) both $f_{\nu}$ and $N_{\rm{eff}}$ allowed to vary. We report the $95\%$ upper limit on $M_{\nu}$, and the mean and $68\%$ confidence interval about the mean for $N_{\rm{eff}}$.}
\begin{ruledtabular}
\begin{tabular}{ccccc}
~ & 1 & 2 & 3 & 4\\
\hline
$M_{\nu}$ (eV) & -- & $< 0.45 $ & -- & $< 0.67 $ \\
$N_{\rm{eff}}$ & -- & -- & $ 3.56\pm 0.32 $ & $ 3.71\pm 0.35 $ \\
 \end{tabular}
\end{ruledtabular}
\label{table:fnunnu}
\end{table}
%****************************************

We find that allowing these small neutrino masses does not significantly change any of our parameter constraints, whereas allowing additional sterile neutrino species does. In particular, we see the physical dark matter density $\Omega_{\rm{DM}}h^2$ is increased while the power spectrum of primordial fluctuations gains an enhancement in both the tilt $n_s$ and amplitude $A_s$ \cite{Bashinsky:2003tk}.

\begin{figure}
\includegraphics[width=8.75cm]{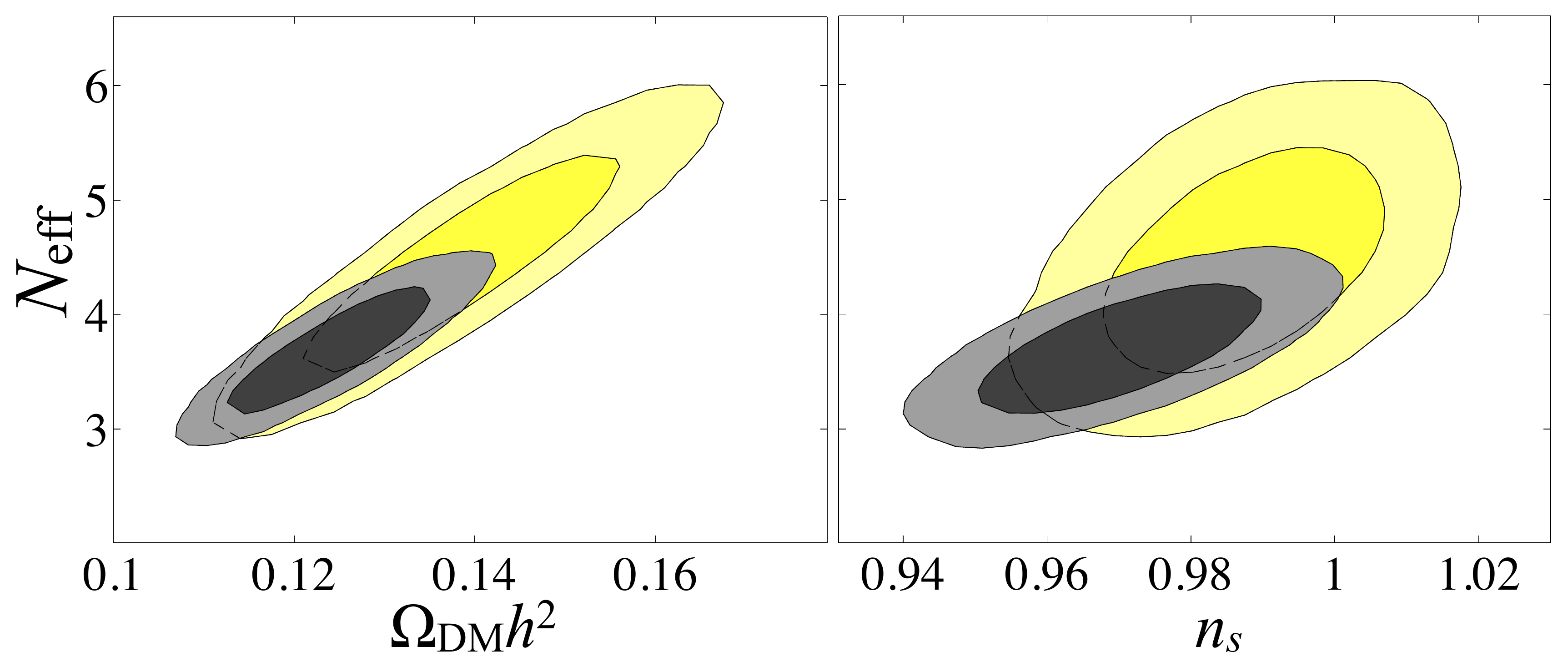}
\caption{Constraints in the $\Omega_{\rm{DM}}h^2-N_{\rm{eff}}$ (left panel) and $n_s-N_{\rm{eff}}$ (right panel) planes without (yellow) and with (grey) SPT CMB data~\cite{Story:2012wx}, for the case where both $N_{\rm{eff}}$ and $f_{\nu}$ are varied.}
\label{fig:degen} 
\end{figure}

In Fig.~\ref{fig:degen} we show how the degeneracies between $N_{\rm{eff}}$ and $\Omega_{\rm{DM}}h^2$ or $n_s$ tighten with the addition of the SPT CMB data, which provides several more peaks in the small-scale regime. As has been noted elsewhere, e.g.~\cite{Giusarma:2012ph}, we find that the inclusion of an $H_0$ prior strengthens our constraints on $N_{\rm{eff}}$. However, we still get meaningful results without it due to our inclusion of SPT data. The damping scale test provides constraints which are independent of low-redshift dynamics, and therefore the specifics of the dark energy model.

%****************************************
%****************************************

%****************************************
\begin{figure*}
\includegraphics[width=\picwidth]{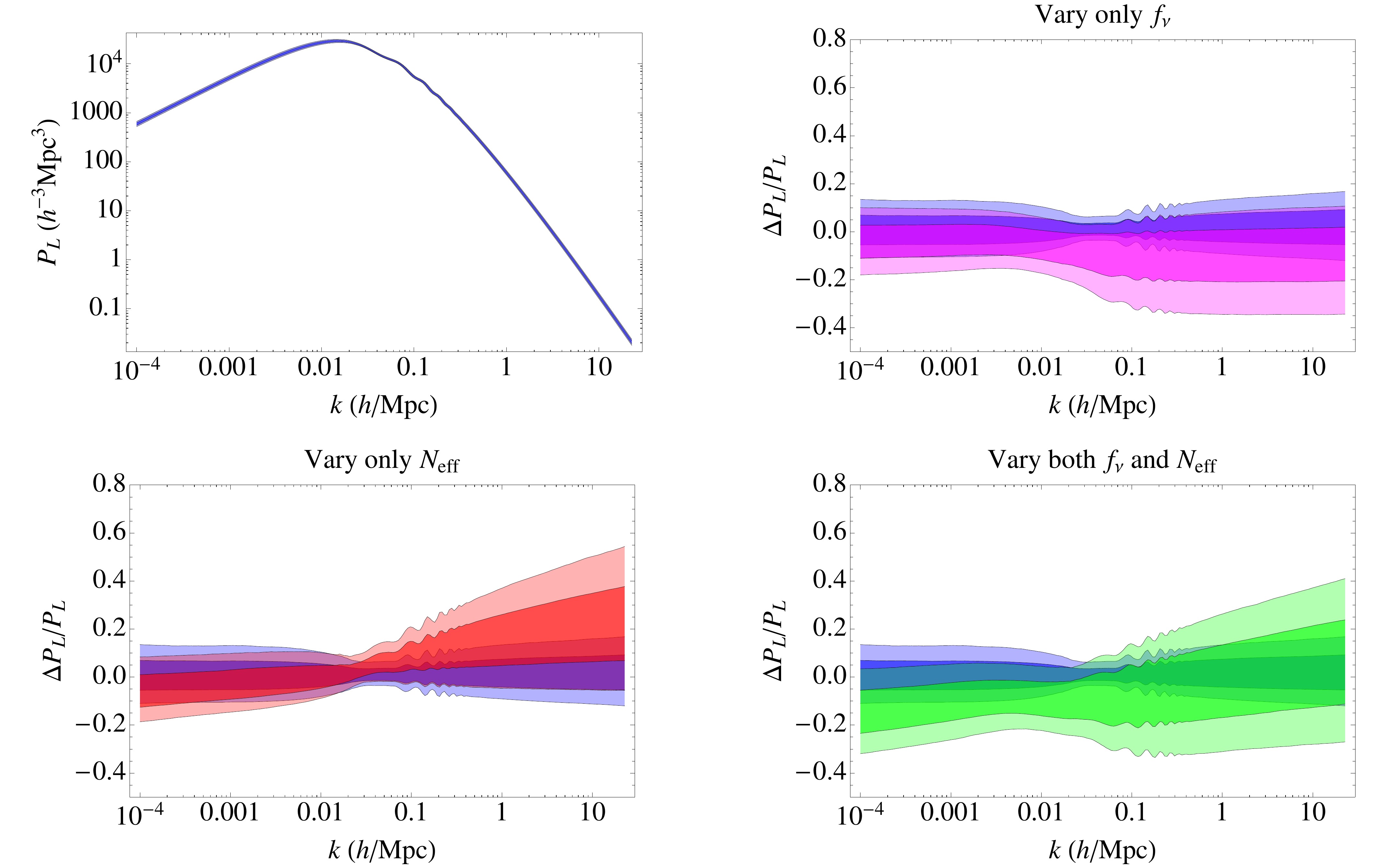}
\caption{Flat $\Lambda$CDM predictions for the $z=0$ linear matter power spectrum, with $k$ in units of $h$ Mpc$^{-1}$ and $P(k)$ in units of $h^{-3}$ Mpc$^{3}$, and the color-coding as before for the four cases: (1) three massless neutrinos (blue); (2) $f_{\nu}$ allowed to vary (magenta); (3) $N_{\rm{eff}}$ allowed to vary (red); and (4) both $f_{\nu}$ and $N_{\rm{eff}}$ allowed to vary (green). The top-right and bottom panels are all plotted with respect to the maximum likelihood ``blue" model prediction, showing the 68\% and 95\% confidence level regions, and with the same axis scales for comparison.}
\label{fig:pk} 
\end{figure*}
%****************************************

In Fig.~\ref{fig:pk} we show how the $z=0$ linear matter power spectrum prediction contours shift with the addition of massive neutrinos, additional neutrino species, or both. The top-left panel shows the baseline (i.e.~three massless neutrinos) prediction, and the top-right and bottom panels are plotted with respect to the maximum likelihood baseline model prediction, with color-coding as before. For plotting purposes we follow the usual convention of taking $P(k)=(2\pi^2/k^3)\Delta^2(k)$, with $k$ in units of $h$ Mpc$^{-1}$. We see that endowing neutrinos with mass serves to suppress structure growth, in accordance with conventional wisdom, despite the similar parameter predictions as seen in Fig.~\ref{fig:params}. We further see that allowing for additional sterile neutrino species serves to enhance structure  on small scales. This is because of an $N_{\rm{eff}}-n_s$ degeneracy.  
A larger $N_{\rm eff}$ suppresses power in the high $\ell$ CMB spectrum due to damping which then allows a compensating increase in $n_s$ or the high-$k$ primordial power spectrum.
Given the preference for additional species seen in Table~\ref{table:fnunnu}, we find that this significantly shifts our $P_L$ contours up for large $k$.

%****************************************
\begin{figure*}
\includegraphics[width=\picwidth]{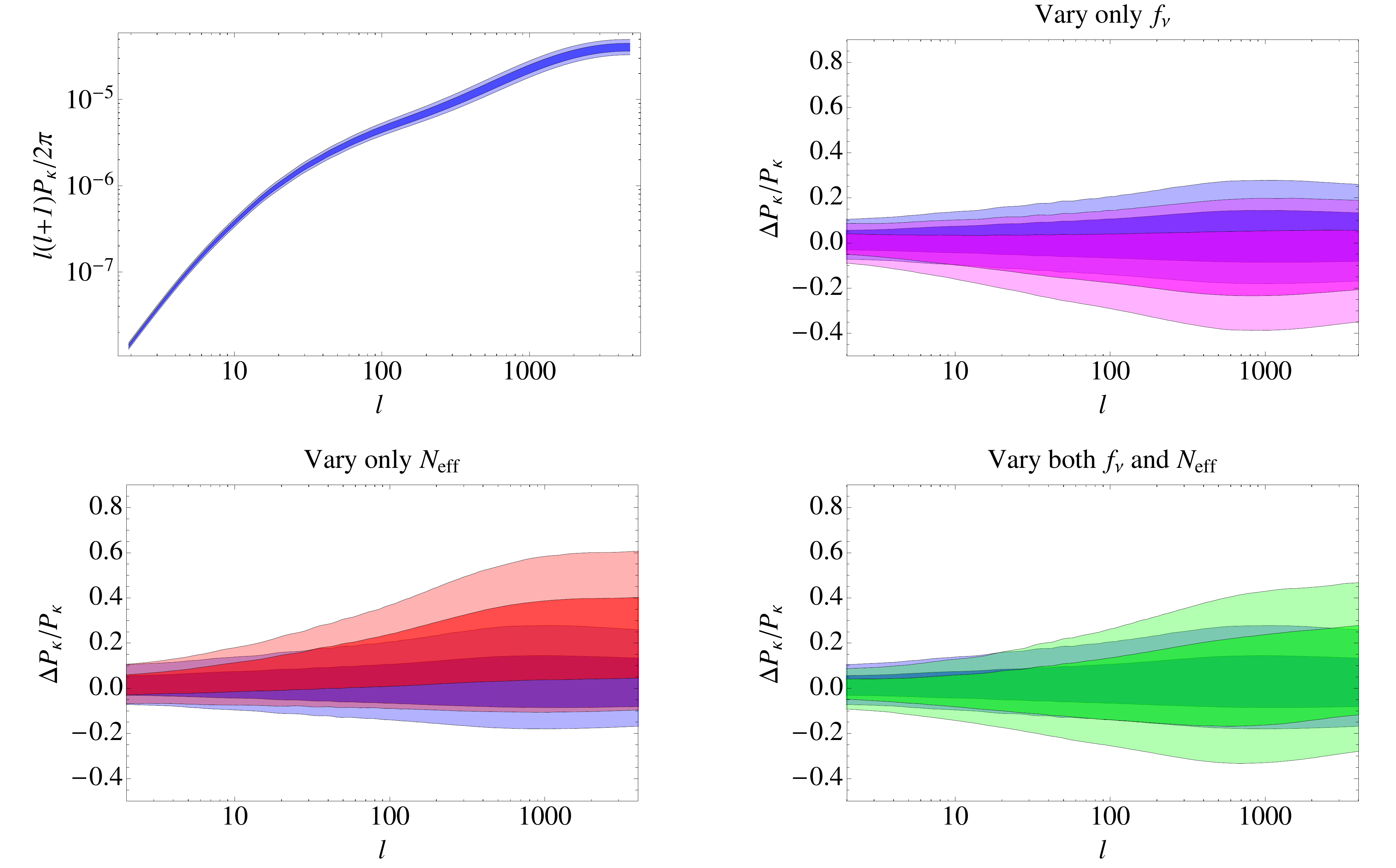}
\caption{Flat $\Lambda$CDM predictions for the ground-based cosmic shear power spectrum, with the color-coding as before for the four cases: (1) three massless neutrinos (blue); (2) $f_{\nu}$ allowed to vary (magenta); (3) $N_{\rm{eff}}$ allowed to vary (red); and (4) both $f_{\nu}$ and $N_{\rm{eff}}$ allowed to vary (green). The top right and bottom panels are all plotted with respect to the maximum likelihood ``blue" model prediction, showing the 68\% and 95\% confidence level regions, and with the same axis scales for comparison.}
\label{fig:shear} 
\end{figure*}
%****************************************

We show the resulting 2D cosmic shear power spectra in Fig.~\ref{fig:shear} for our model ground-based weak lensing survey, again with  the top-left panel showing the baseline prediction, and the top-right and bottom panels plotted with respect to the maximum likelihood baseline model prediction. 

\begin{figure}
\includegraphics[width=8cm]{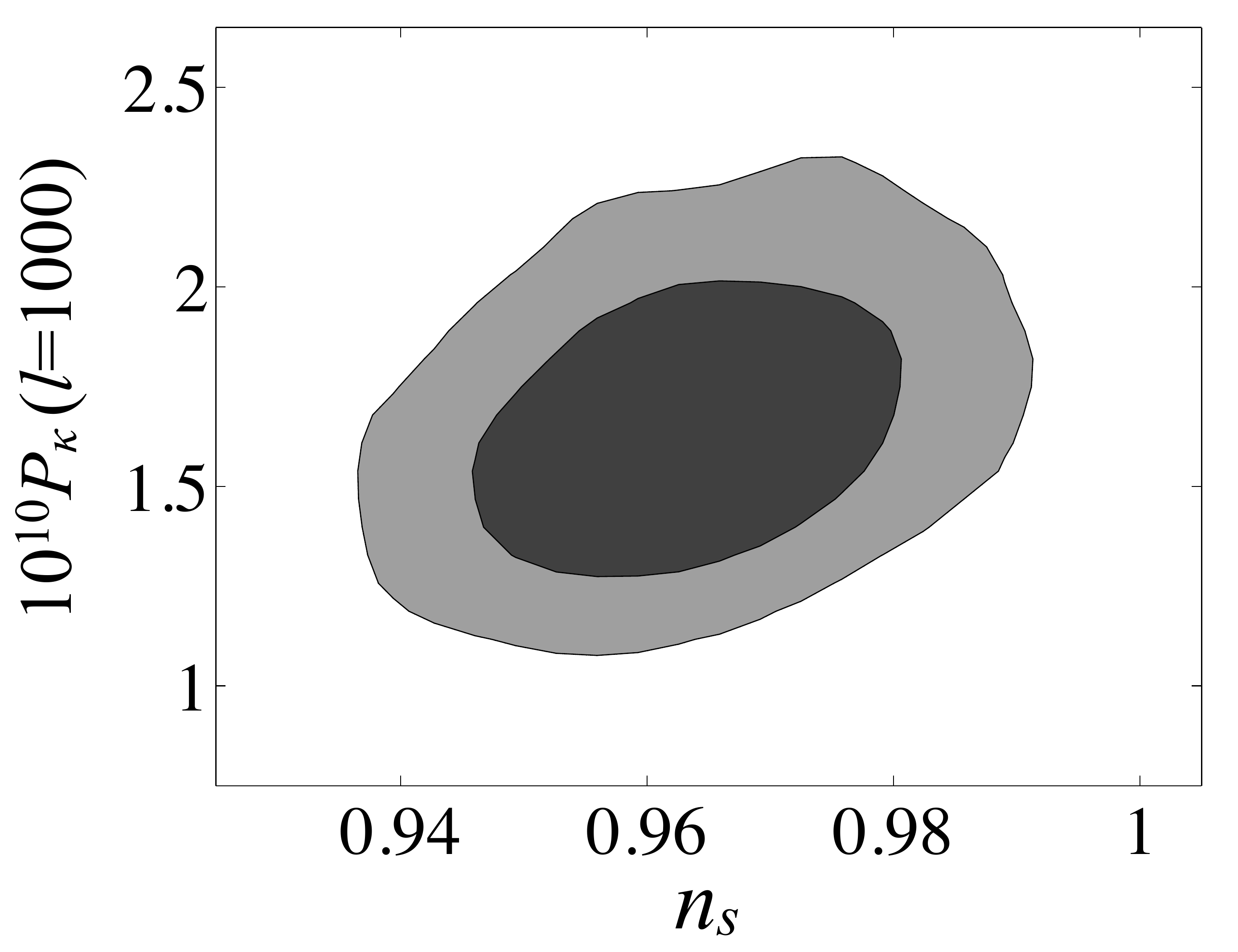}
\caption{Constraints on the ground-based cosmic shear power spectrum amplitude at $l = 1000$ (multiplied by $10^{10}$) vs $n_s$, for the case where $N_{\rm{eff}}$ is varied but $f_{\nu}=0$, showing the 68\% and 95\% contours.}
\label{fig:tilt} 
\end{figure}

In the context of the flat $\Lambda$CDM model, an observed deficit of small-scale
cosmic shear of between $20-40\%$ would indicate finite neutrino mass and could
not be explained by other currently allowed cosmological parameter variations.
An observed excess of $20-60\%$ would indicate extra neutrino species and 
comes from the freedom to raise the tilt due to the $N_{\rm eff}-n_s$ degeneracy
in the CMB.    Cosmic shear measurements provide a means of breaking this degeneracy. 
We illustrate the issue in Fig.~\ref{fig:tilt}, where we plot our constraints on the cosmic shear power at $l = 1000$ vs $n_s$ for the case where $N_{\rm{eff}}$ is varied but $f_{\nu}=0$.
The case where both $f_\nu$ and $N_{\rm eff}$ are allowed
to vary is harder to distinguish in that the two effects can partially compensate
for each other.   On the other hand, a further breaking of the  $N_{\rm eff}-n_s$ 
degeneracy is expected from the {\it Planck} survey \cite{Planck}, thereby allowing
these mixed cases to be better separated with cosmic shear.

%****************************************
%****************************************

\begin{figure}
\includegraphics[width=8cm]{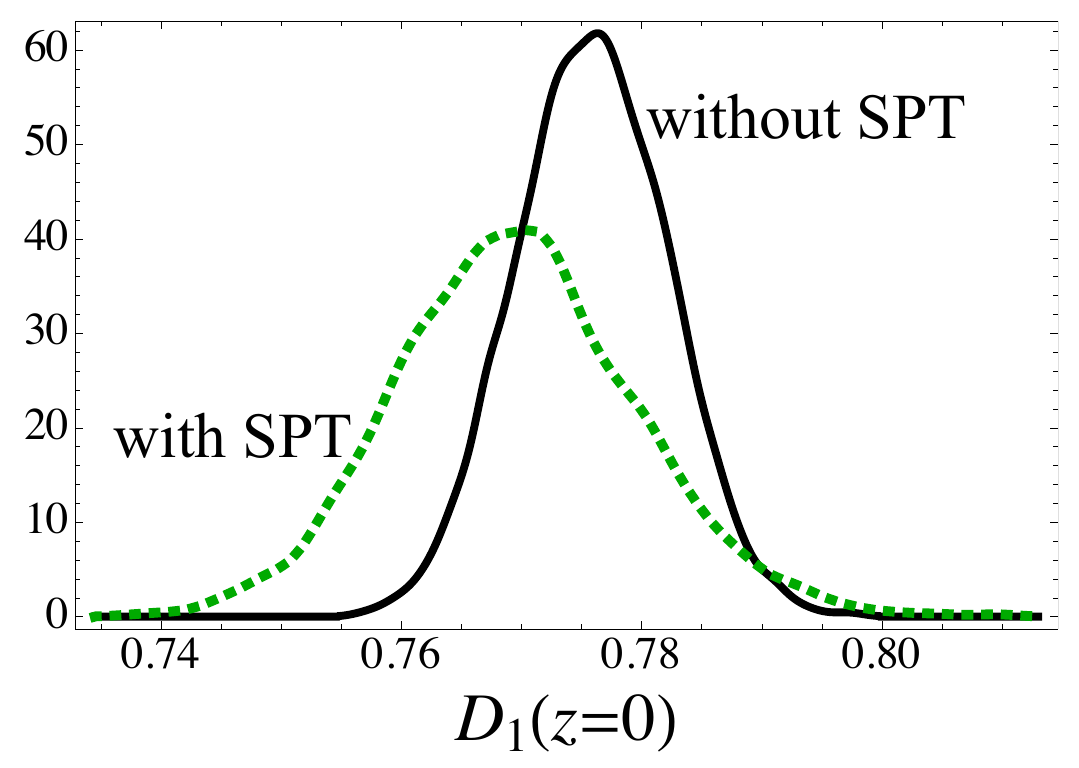}
\caption{One-dimensional constraints on the scale-independent growth function $D_{1}$ at $z = 0$, for the case where both $N_{\rm{eff}}$ and $f_{\nu}$ are allowed to vary, without (black solid) and with (green dotted) the addition of SPT CMB data.  Note that the additional data shifts and weakens the joint constraint, thereby indicating
tension in the data sets.}
\label{fig:growth} 
\end{figure}

It is interesting to note that while the addition of SPT data tightens the predictions
for the shear power spectrum it actually weakens and shifts the predictions on the growth
function, as we show in Fig.~\ref{fig:growth} for $z=0$.  This reflects a mild tension between this data set and the BAO measurements caused by its improved measurement of $\Omega_{\rm DM} h^2$. 
Unlike the similar tension between BAO and $H_0$ for the flat $\Lambda$CDM model, this
tension is not alleviated by allowing $N_{\rm eff}$ (or $f_\nu$) to vary. 
On the other hand, constraints on the growth function are not the dominant source
of error for shear predictions and so this tension is not relevant for our purposes.  

%****************************************
%****************************************

Beyond the flat $\Lambda$CDM model, there are other possibilities that can
explain a deficit or excess of small-scale shear.  
When the dark energy equation of state is generalized to allow quintessence, only a deficit can arise due to the restriction that $w\ge -1$ \cite{Vanderveld:2012ec}.
Hence these cases can masquerade as massive neutrino models unless
further information on the shape and redshift dependence of the power spectrum
is obtained.    Without neutrino number changes, an excess cannot be explained
by quintessence and hence would indicate more exotic cosmic acceleration
physics with enhanced forces in the dark sector.  Again, the expected 
improvements from the {\it Planck} survey will help distinguish between these
possibilities.

Likewise we have also tested the robustness of these results to dropping the flatness assumption.  Allowing for curvature significantly degrades our constraints on $M_{\nu}$, where we find the $95\%$ limits expand to $M_{\nu}<1.68$ eV and $M_{\nu}<1.88$ eV for the cases when $f_{\nu}$ only is varied and for when both $f_{\nu}$ and $N_{\rm eff}$ are varied, respectively. On the other hand, our constraints on $N_{\rm eff}$ are not significantly different then those of the flat case. We further find that our high-redshift growth function constraints are weakened, but since the confidence contours are still well within the $1\%$ range we find that there is not a significant effect on the shear predictions. Indeed our shear predictions are qualitatively the same.

%%%%%%%%%%%%%%%%%%%%%%%%%%%%%%%%%%%%%%%%%%%%%%%%%%%%%%%%
\section{Discussion}
\label{sec:discussion}

Inspired by recent evidence for massive neutrinos and the possibility of additional species, we have provided an analysis of the signatures of such ``nonstandard" neutrino physics on the weak lensing shear power spectrum. By using observations of the expansion history from Type Ia SNe, BAO, and local measures of $H_0$, coupled with observations of the CMB, we can predict future structure-growth observables such as those from weak lensing. From doing so for our four different scenarios -- the standard case with three massless neutrinos, three massive neutrinos, any number of massless neutrinos, and three massive neutrinos with any number of massless neutrinos -- we can look for signatures that cannot be mimicked by
any currently allowed  variation in the other parameters of the flat $\Lambda$CDM
model.  We present only results for a representative model ground-based survey here, but the results for any survey configuration will be qualitatively the same.

For our most general case, where we vary both neutrino mass and number of species, we find parameter constraints that are consistent with the current literature. Using only distance measures and the CMB, we are able to constrain $M_{\nu} < 0.67$ eV at the 95\% confidence level and $N_{\rm{eff}} = 3.71\pm 0.35$.  

Such variations from the standard neutrino parameters allow changes
in the predicted shear power spectrum that cannot be mimicked by other flat $\Lambda$CDM parameters.   For example shear deficit in the 20-40\% range would indicate neutrino masses near saturation of current bounds with $N_{\rm{eff}} \sim 3$ 
whereas shear excess in the 20-60\% range would indicate extra neutrino 
species and lower masses.  The latter is because of a partial 
degeneracy between raising $N_{\rm{eff}}$ and spectral tilt of the primordial power spectrum $n_s$ (see Fig.~\ref{fig:tilt}).

It has been noted previously that generalizing from the flat $\Lambda$CDM model to quintessence can only serve to reduce the matter power spectrum \cite{Mortonson:2010mj,Vanderveld:2012ec}, within the context of the standard (three massless) neutrino scenario. 
Quintessence effects can therefore mimic the deficit predicted by massive
neutrinos without further information from the shape and redshift dependence
of the power spectrum.

A future measurement of excess cosmic shear could provide
supporting evidence for extra neutrino species.  An excess could also be explained by
more exotic acceleration physics that enhances structure growth.  Since the neutrino effect
comes from the primordial power spectrum, improved measurements from the {\it Planck} satellite would help distinguish these options should an excess be found.

%%%%%%%%%%%%%%%%%%%%%%%%%%%%%%%%%%%%%%%%%%%%%%%%%%%%%%%
%%%%%%%%%%%%%%%%%%%%%%%%%%%%%%%%%%%%%%%%%%%%%%%%%%%%%%%%
\begin{acknowledgments}

We thank Bradford Benson, Michael Mortonson, and Kyle Story for useful conversations, and Antonio Cuesta for assistance with BAO data. RAV and WH acknowledge the support of the Kavli Institute for Cosmological Physics at the University of Chicago through grants NSF PHY-0114422 and NSF PHY-0551142 and an endowment from the Kavli Foundation and its founder Fred Kavli. WH acknowledges additional support from DOE contract DE-FG02-90ER-40560 and the Packard Foundation.

\end{acknowledgments}

\bibliography{neutrinos}

\end{document}